\documentclass[aps,prl,twocolumn,groupedaddress,floatfix,preprintnumbers]{revtex4}
\pdfoutput=1

\usepackage[dvips]{color}
\usepackage{latexsym,amsmath,amsfonts,amssymb}
\usepackage{graphicx}
\usepackage{FBcommands}

\begin{document}

\preprint{PUPT-2341}

\title{Holographic Fermi arcs and a $d$-wave gap}


\author{Francesco Benini}
\author{Christopher P. Herzog}
\author{Amos Yarom}
\affiliation{Princeton University, Princeton, NJ 08542}


\date{\today}

\begin{abstract}

We study fermion correlators in a holographic superfluid with a $d$-wave (spin two) order parameter. We find that, with a suitable bulk Majorana coupling, the Fermi surface is anisotropically gapped. At low temperatures the gap shrinks to four nodal points. At high temperatures the Fermi surface is partially gapped generating four Fermi arcs.

\end{abstract}


\maketitle

\textit{Introduction and Summary.} --
Gauge gravity duality \cite{Maldacena:1997re,Gubser:1998bc,Witten:1998qj}
provides a useful tool for understanding the strongly coupled planar limit of gauge theories.
At low temperatures and non-zero density these
theories exhibit properties similar to the ones observed in condensed matter systems.
Using gauge gravity duality, various authors have constructed configurations which exhibit
superconductivity \cite{Gubser:2008px,Hartnoll:2008vx}, a $p$-wave order parameter \cite{Gubser:2008zu,Gubser:2008wv,Roberts:2008ns}, and nodes in fermion correlators \cite{Gubser:2010dm,Ammon:2010pg}. For reviews see \eg{} \cite{Hartnoll:2009sz,Herzog:2009xv,Horowitz:2010gk}.

\begin{figure}[b]
\vspace{-.6em}
\begin{center}
\includegraphics[height=4.3cm]{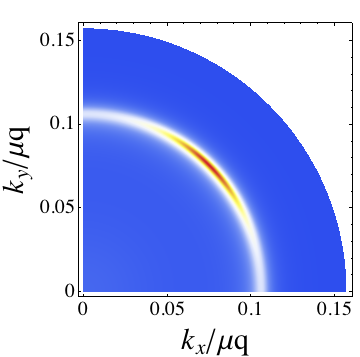}
\hspace{\stretch{1}}
\includegraphics[height=4.3cm]{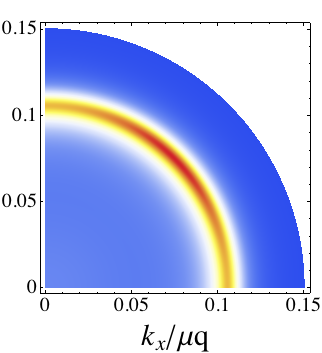}

\caption{(Color online) A density plot of the fermion spectral function
$\rho$ defined in (\ref{E:spectral})
evaluated at $\omega=0$ for temperatures $T=0.49 \, T_c$ (left) and $T=0.59 \, T_c$ (right). Red and blue correspond to large and small values of the spectral function.\label{fig: density plot MDC}}
\vspace{-1.6em}
\end{center}
\end{figure}

In this work we pursue the analogy with condensed matter systems further and study fermion correlators in large $N$ gauge theories with a $d$-wave (spin two) order parameter. Our main finding is that in the phase where the spin two field condenses, and at non-zero temperature, the spectral function for the fermions develops asymmetric features including an angle dependent gap along the Fermi surface (FS). These features are exhibited in fig.~\ref{fig: density plot MDC}.
For particular values of the coupling between the spin two field and the fermions, this gap vanishes along a finite interval reminiscent of the ``Fermi arcs'' observed in the pseudogap region of certain high $T_c$ superconductors \cite{Norman:1998,Kanigel:2006,Kanigel:2007}.
We will discuss similarities and differences between our results
and the experimental observations
in high $T_c$ materials towards the end of this work.

\textit{Setup.} --
Gauge gravity duality relates the partition function of a gravitational theory with an asymptotically
anti-de-Sitter (AdS) solution to the partition function of a conformal field theory (CFT) in one spatial dimension less. We will refer to the gravitational theory as the bulk theory and to its CFT dual as the boundary theory.

The bulk fields of a holographic superconductor with a $d$-wave order parameter in $3+1$ dimensions are the gravitational field, a $U(1)$ field and a massive charged (symmetric) spin two field $\varphi_{\mu\nu}$
\cite{Chen:2010mk,Herzog:2010vz,Ours:2010}.
The unique matter Lagrangian quadratic in $\varphi_{\mu\nu}$, containing operators of dimension four or less, and describing the correct number of propagating degrees of freedom is
 \cite{Velo:1972rt,Buchbinder:1999ar}
\begin{multline}
\label{E:spin2lag}
\cL =
- \frac{1}{4}F_{\mu\nu}F^{\mu\nu} - i q_\varphi F_{\mu\nu} \varphi^{\mu\rho*} \varphi^\nu_\rho \\
- |D_\rho \varphi_{\mu\nu}|^2 + 2 |\varphi_\mu|^2 + |D_\mu \varphi|^2 - \big[ \varphi^{\mu*} D_\mu \varphi + \text{h.c.} \big] \\
- m_\varphi^2 \big( |\varphi_{\mu\nu}|^2 - |\varphi|^2 \big) + 2 R_{\mu\rho\nu\lambda} \varphi^{\mu\nu*} \varphi^{\rho\lambda} - \frac R4 |\varphi|^2
\end{multline}
together with the constraint that the metric is Einstein.
We introduced
the notation $\varphi_\mu \equiv D^\nu \varphi_{\nu\mu}$ and $\varphi \equiv \varphi^\nu_\nu$. The $U(1)$ gauge field is denoted $A$ and its field strength is
$F=dA$. The Ricci scalar and Riemann tensor are $R$ and $R_{\mu\nu\rho\sigma}$ respectively. The fully covariant derivative $D_\mu$ is related to the metric covariant derivative $\nabla_\mu$ 
through $D_\mu = \nabla_\mu - i q_\varphi A_\mu$.

The Lagrangian (\ref{E:spin2lag}) should be treated using the language of effective field theory.
The equations of motion (EOMs) following from $\cL$ exhibit non-causal
behavior that can be corrected by adding higher order terms in $F$ and its gradients.
See \cite{Ours:2010} for a detailed discussion.

The Einstein condition on the metric restricts us to the probe limit, where the charge of the spin two field is large enough so that the matter fields do not backreact on the metric.
Thus, the solution to the Einstein equations is the AdS$_4$ Schwarzschild black hole whose line element is
\begin{equation}
ds^2 =( -f \, dt^2 + dx^2 + dy^2 + dz^2 / f  ) \, L^2 / z^2
\end{equation}
where
$ f(z) = 1 - \left( 4 \pi T z/3 \right)^3 $ and $T$ is the Hawking temperature of the black hole and also the temperature of the field theory.
To solve for the matter fields we consider an
ansatz where only
\begin{equation}
\label{E:ansatz}
A_t(z) \;, \qquad \varphi_{xy}(z) \;,\qquad \varphi_{xx} = - \varphi_{yy} \equiv \varphi_\Delta(z)
\end{equation}
are non-zero and $\varphi_{\mu\nu}$ is real. We note that with this ansatz, $\varphi = \varphi_\mu = F_{\mu\rho} \varphi^\rho_\nu = 0$. Under a spatial rotation of an angle $\theta$ in the $(x,y)$-plane, the spin two field transforms as
\begin{equation}
\label{E:rotation}
\mat{ \varphi_\Delta \\ \varphi_{xy} } \;\xrightarrow{R(\theta)}\; \mat{ \cos2\theta & -\sin2\theta \\ \sin2\theta & \cos2\theta} \, \mat{ \varphi_\Delta \\ \varphi_{xy} } \;.
\end{equation}

In \cite{Ours:2010} it was noted that by redefining $\varphi_{\mu\nu} = z^2 \tilde\varphi_{\mu\nu}$ and choosing an angle $\theta$ at which either $\varphi_{\Delta}$ or $\varphi_{xy}$ vanishes, the EOMs
and boundary conditions (BCs) for $A_t$ and $\tilde{\varphi}$ reduce to those of the Abelian Higgs theory first discussed in \cite{Gubser:2008px, Hartnoll:2008vx}. This identification guarantees that when the chemical potential $\mu$ of the boundary theory does not vanish, there exists a critical temperature $T_c$ below which the spin two field condenses.

We are after
the spectral function $\rho(\omega, \vk)$ for fermion operators $\cO$ in the condensed phase of the boundary theory. A study of thermal fermion correlators in gauge gravity duality was initiated in
\cite{Iqbal:2009fd,Lee:2008xf,Liu:2009dm,Cubrovic:2009ye}
and we will follow their approach. Consider the bulk fermion action
\begin{multline}
\label{E:fermilag}
\cL_\Psi = i \oPsi \big( \Gamma^\mu D_\mu - m \big) \Psi \\
+ \eta^* \varphi_{\mu\nu}^* \, \overline{\Psi^c} \Gamma^\mu D^\nu \Psi - \eta \, \oPsi \Gamma^\mu D^\nu (\varphi_{\mu\nu} \Psi^c) \;,
\end{multline}
where now $D_\mu = \partial_\mu + \frac14 w_{\mu,\ulambda\usigma} \Gamma^{\ulambda\usigma} - i q A_\mu$ with $w$ the spin connection and $\Psi$ a Dirac spinor.
Gauge invariance of the coupling between the fermions and the spin two field dictates that
the charge of $\Psi$ is
$q = q_\varphi/2$.
The bulk gamma matrices $\Gamma^{\mu}$ satisfy $\{\Gamma^{\mu},\Gamma^{\nu}\}=2g^{\mu\nu}$.  The conjugate spinor $\Psi^c$ is defined by $\Psi^c \equiv C \Gamma^{\ut} \Psi^*$, where $C\Gamma^\mu C^{-1} = - \Gamma^{\mu\trans}$. Vielbein indices are underlined. The phase of the coupling constant $\eta$ can be absorbed in a redefinition of $\Psi$: we will use this freedom to set $\eta \geq 0$.

If we restrict ourselves to couplings of mass dimension smaller than six then with the ansatz \eqref{E:ansatz} and the choice $q_\varphi=2q$,
the only other possible coupling between the spin two field and the fermions is $|\varphi_{\mu\nu}|^2 \oPsi \left(c_1 + c_2\Gamma^5\right)\Psi$. The latter provides an effective mass for the fermions and does not contribute to the asymmetric features of the spectral function which we would like to exhibit. Therefore we will not consider it further. Note that had one chosen $q = q_\varphi$, an interaction term of the form $\varphi_{\mu\nu}^* \varphi^{\mu\nu*} \overline{\Psi^c} \left(c_3 + c_4 \Gamma^5 \right) \Psi$ would have been possible, leading to an $s$-wave instead of a $d$-wave gap.

The Dirac equation following from (\ref{E:fermilag}) is given by
\be
\label{Dirac equation}
0 = \big( \Gamma^\mu D_\mu - m \big) \Psi + 2i\eta \varphi_{\mu\nu} \Gamma^\mu D^\nu \Psi^c + i \eta \varphi_\mu \Gamma^\mu \Psi^c \;.
\ee
We find it convenient to introduce a rescaled spinor $\psi = (-g \cdot g^{zz})^{1/4} \Psi$.
Since the Dirac equation couples $\Psi$ and $\Psi^c$, we consider a solution of the form
\be
\label{Fansatz}
\psi = e^{-i\omega t + i \vec k\cdot \vec x} \, \zeta(z) + e^{i\omega t - i \vec k\cdot \vec x} \, \tilde{\zeta}(z) \;,
\ee
where $\zeta$ and $\ttilde{\zeta}$ can be decomposed into two two-component spinors: $\zeta = (\zeta_1, \zeta_2)^\trans$. To write equation \eqref{Dirac equation} explicitly we must choose a particular representation for our spinors. Our bulk gamma matrices are given by
\bea
\Gamma^{\ut} &= \left( \begin{smallmatrix} - i\sigma_2 & 0 \\ 0 & i \sigma_2 \end{smallmatrix} \right) \quad &
\Gamma^{\uz} &= \left( \begin{smallmatrix} \sigma_3 & 0 \\ 0 & \sigma_3 \end{smallmatrix} \right) \\
\Gamma^{\ux} &= \left( \begin{smallmatrix} \sigma_1 & 0 \\ 0 & \sigma_1 \end{smallmatrix} \right) &
\Gamma^{\uy} &= \left( \begin{smallmatrix} 0 & -i\sigma_2 \\ i\sigma_2 & 0 \end{smallmatrix} \right) \,.
\label{E:gammas}
\eea
In this representation $C = \Gamma^{\ut}$.
The boundary gamma matrices are defined by the action of $\Gamma^{\ulambda\usigma}$ on the positive eigenspace of $\Gamma^{\uz}$:
$\gamma^t = -i\sigma_2$, $\gamma^x = \sigma_1$, $\gamma^y = -\sigma_3$.
Given (\ref{E:gammas}), the Dirac equation \eqref{Dirac equation} reduces to
\bea
\label{E:Deqns}
0 &= D_{(1)} \, \zeta_1  + 2\eta (g^{xx})^{\frac32} k_x \big[ \varphi_\Delta \sigma_1 \tilde{\zeta}_1^{*}  -i \varphi_{xy} \sigma_2 \tilde{\zeta}_2^{*} \big] \\
0 &= D_{(2)} \, \zeta_2  + 2\eta (g^{xx})^{\frac32} k_x \big[ \varphi_\Delta \sigma_1 \tilde{\zeta}_2^{*} +i \varphi_{xy} \sigma_2 \tilde{\zeta}_1^{*} \big]
\eea
where we have set $k_y=0$.
Arbitrary $\vk$ can be recovered using (\ref{E:rotation}).
The differential operators $D_{(\alpha)}$ are defined as
\begin{multline}
D_{(\alpha)}
= \sqrt{g^{zz}} \, \sigma_3 \partial_z - m \\
+(-1)^{\alpha} (\omega + qA_t) \sqrt{-g^{tt}}\, \sigma_2 + ik_x \sqrt{g^{xx}} \, \sigma_1 \;.
\end{multline}
The equations for $\ttilde{\zeta}_\alpha$ can be obtained from \eqref{E:Deqns} with the substitution $(\omega,k_x) \to (-\omega,-k_x)$ and $\zeta \to \ttilde{\zeta}$.

Once we find a solution to the Dirac equation in the bulk, we can use it to determine correlators of fermion operators $\cO$ in the boundary theory. In particular, we will be interested in the Fourier transform of the retarded Green's function  $G_R(t,\vec x) = i \Theta(t) \vev{ \{\cO(t,\vec x) , \cO^\dag(0) \} }$, from which we can compute the spectral function.
In what follows we summarize the procedure for obtaining the boundary theory Green's function. The reader interested in a detailed exposition is referred to \cite{Iqbal:2009fd}.

Consider the near boundary (small $z$) series expansion of the fermion fields,
\be
\zeta_\alpha \,\stackrel{z\to 0}{\approx}\, \mat{ R_\alpha \\ O(z)} z^{Lm} + \mat{ O(z) \\ (\sigma_1 S)_\alpha } z^{-Lm}
\ee
where the index $\alpha = 1,2$ is a spinor index. The component $S_\alpha$ acts as a source term for the fermion boundary operator $\cO_{\alpha}$ and $R_\alpha = \langle \cO_{\alpha} \rangle$. A similar expression involving $\ttilde{R}_\alpha$ and $\ttilde{S}_\alpha$ follows from a series expansion of \ttilde{\zeta}.
Since the equations of motion are linear, the two-component spinor $R$ is linearly related to $S$ and \ttilde{S} \footnote{On the boundary $C = - \gamma^t$, so $S^c = C\gamma^t S^* = S^*$.},
\be
R_\alpha = \cM\du{\alpha}{\beta} S_\beta + \tilde \cM\du{\alpha}{\beta} \tilde S_\beta^* \;.
\ee
The prescription for computing the retarded Green's function $G_R(\omega, \vk)$ is
\be
\label{Green's function formula}
G_R = -i \cM \gamma^t
\ee
where the EOMs \eqref{E:Deqns} should be solved with BCs which are infalling at the horizon.
With a slight abuse of terminology, we define the spectral function $\rho$ as
\begin{equation}
\label{E:spectral}
\rho = \Tr \Im G_R
\end{equation}
where the trace is over the spin degrees of freedom.

\textit{The appearance of a gap.} --
To compute $\rho$ we need to solve the Dirac equation.
However, even without an explicit solution to \eqref{E:Deqns} we can infer the appearance of an angle dependent gap in $\rho$ whose size vanishes
along four nodal directions.

Equation \eqref{E:Deqns} with $\eta=0$ is identical to the Dirac equation for a free charged fermion in an $s$-wave background.
The Green's function for fermions in the condensed phase of a holographic $s$-wave geometry was studied by the authors of \cite{Chen:2009pt,Faulkner:2009am,Gubser:2009dt} for various values of the charge $q$ and at zero temperature. The salient features of their analysis are that the spectral function $\rho$ has support inside an ``infrared lightcone'' which, in the probe limit, would correspond to the region $\omega \geq |\vk|$. In addition $\rho$ may have non-vanishing support along codimension-one surfaces outside this lightcone. The intersection of these codimension-one surfaces with the $\omega=0$ plane defines Fermi momenta $k_F$.
The location of the codimension-one surfaces coincide with the normal modes of $\zeta_1$ and $\zeta_{2}$: Suppose that at $k_y=0$, $k_x > 0$, $\zeta_1$  has a normal mode which contributes to a pole of $G_{R,11}$. Then, at $k_y=0$, $k_x<0$, $\zeta_2$ will have a normal mode which contributes to a pole of $G_{R,22}$. As expected, the spectral function $\rho$ is rotationally invariant. From (\ref{Fansatz}) it should be clear that the locations of the normal modes of \ttilde{\zeta} can be obtained from those of $\zeta$ with the replacement $(\omega,k_x) \to (-\omega, -k_x)$.

We would like to use the results of \cite{Chen:2009pt,Faulkner:2009am,Gubser:2009dt} as a starting point
to study the distribution of normal modes of the Dirac equation \eqref{E:Deqns} in the presence of a non trivial coupling $\eta$. Unfortunately, the analysis of \cite{Gubser:2009cg, Horowitz:2009ij} implies that, in the probe approximation and with $m_\varphi^2 \neq 0$, one does not have a good zero temperature limit of the spin two condensate. Therefore, in what follows, we will restrict ourselves to finite temperature configurations.
At small but non-vanishing temperature, one expects the normal modes to broaden into quasi-normal modes whose width decreases with decreasing temperature. We have verified this expectation numerically.

Consider first a configuration where $\eta=0$ and $\varphi_\Delta = 0$ and focus on a single normal mode of, say, $\zeta_1$ and the corresponding mode of $\ttilde{\zeta}_2$. These 
modes cross at $\omega=0$ and $k_x = k_F$. Turning on $\eta$ couples $\zeta_1$ and $\ttilde{\zeta}_2$  and eigenvalue repulsion replaces this degeneracy by a gap.
This gap-generating feature of \eqref{E:fermilag} is similar to the mechanism described in \cite{Faulkner:2009am}, where a gap was generated in an $s$-wave superconductor by adding a Majorana-like coupling to the fermion action. Next, consider a configuration where the condensate is rotated by $\pi/4$ (\ie{} $\varphi_{xy}=0$). When $\eta>0$,  $\zeta_1$ and $\ttilde{\zeta}_1$ are coupled. Since their zero modes do not cross along the $\omega=0$ plane a gap is not generated. For other angles one expects a continuous transition from the ungapped point (the node) to the gapped point (the anti-node).
Generically, multiple FSs will be observed for a positive mass spin two condensate. When $
\eta \neq 0$ these multiple FSs will appear as an intricate angular dependent band-like structure.
In what follows we focus on a single FS.
An analysis of the lifting of the degeneracy in the presence of multiple FSs
is left for future work.

\textit{Results.} -- We solved \eqref{Dirac equation} numerically using a shooting algorithm, first
to generate the condensate and then to compute the Green's function. We set $L^2m^2_\varphi = 7/4$ (corresponding to a $d$-wave order parameter of conformal dimension 7/2), $m=0$ and $\eta = 0.02\, q L$. Numerically we found that once $\eta$ becomes too large, there exists a critical temperature below which the spectral function exhibits a gap even at $\theta = \pi/4$.
In this sense, the value of $\eta$ was tuned to be small.
In presenting our results, we set $\varphi_\Delta = 0$ and let $\theta$ be the angle between $\vk$ and the $x$-axis.

\begin{figure}[t]
{\includegraphics[width=.49\columnwidth]{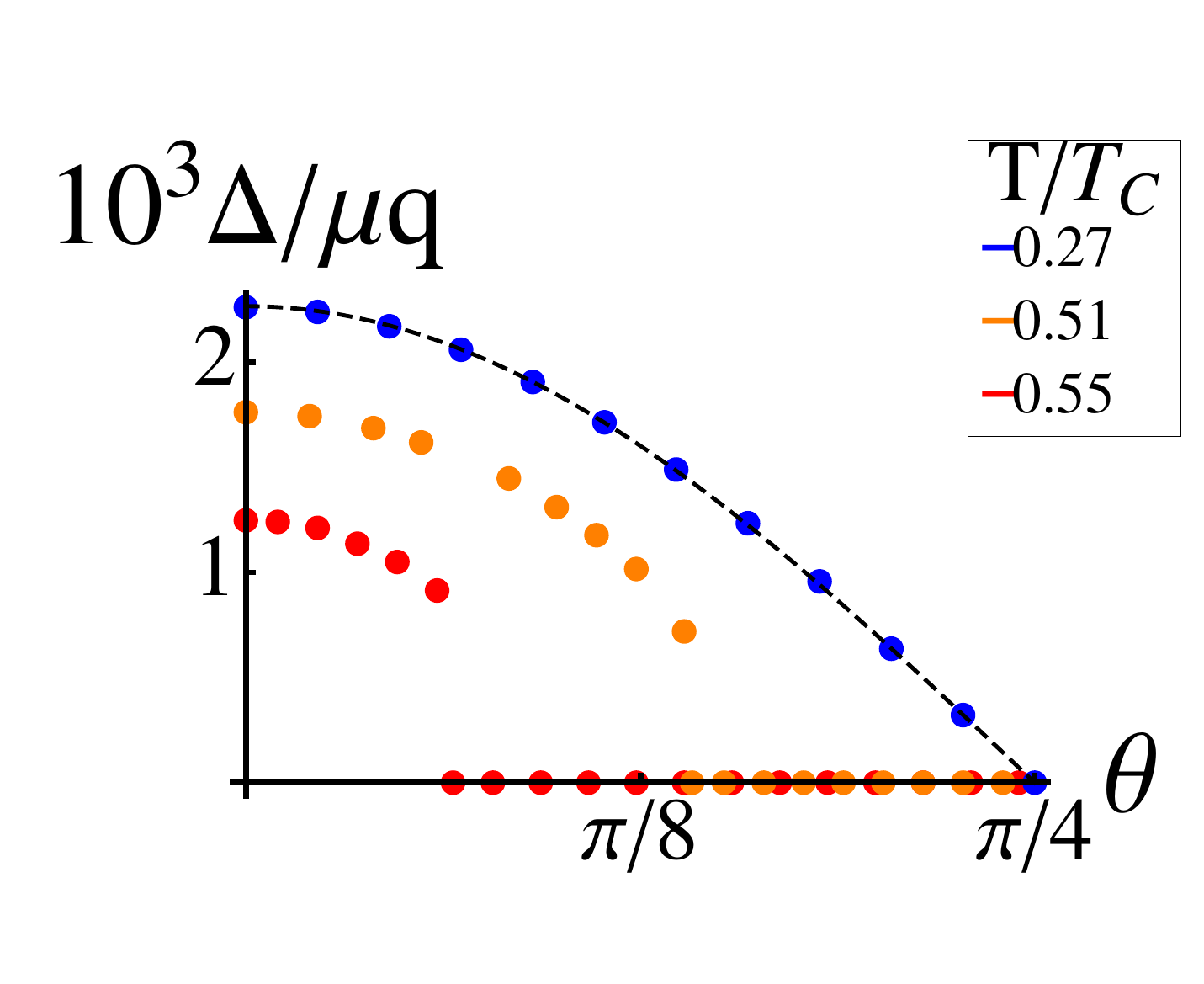}}
\raisebox{0pt}{\includegraphics[width=.49\columnwidth]{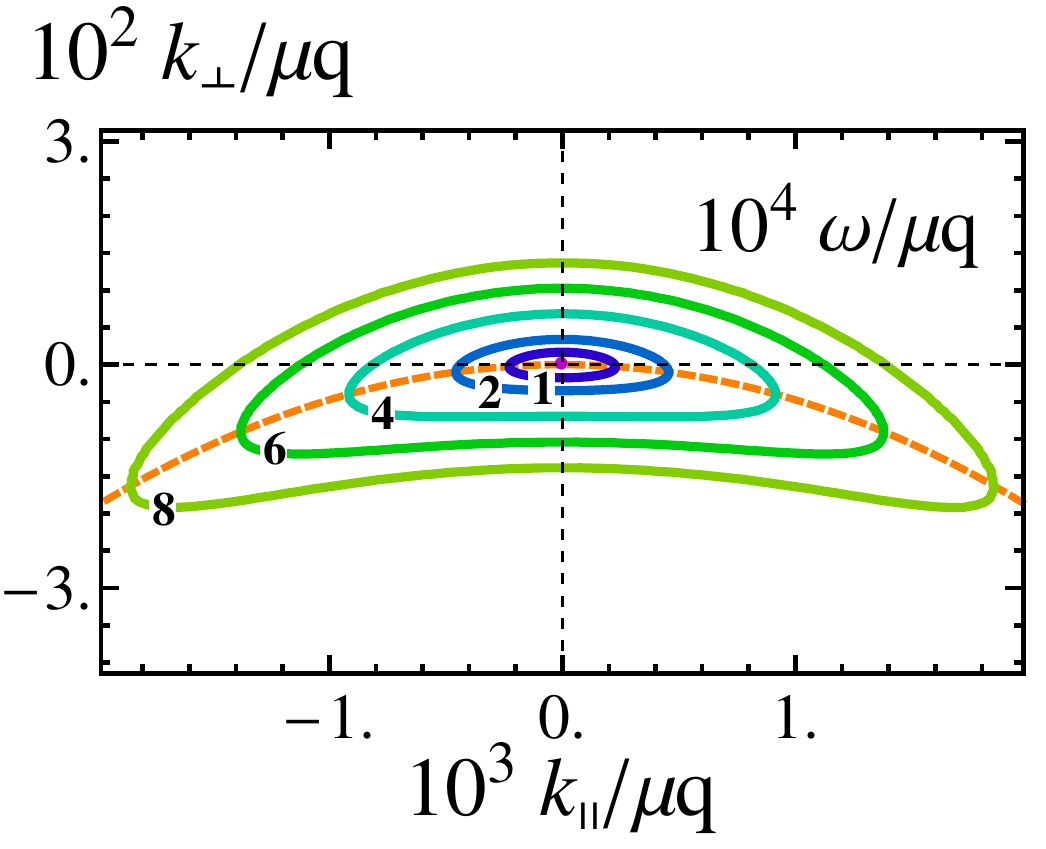}}
\vspace{-2.5em}

\begin{minipage}[t]{.47\columnwidth}
\caption{(Color online) Plots of the gap $\Delta(\theta,T)$ at different temperatures and angles. When the FS is completely gapped, $\Delta$ can be fit by $\Delta_0(T) \cos2\theta$ (dashed). \label{fig: gap vs angle}}
\end{minipage}
\hspace{\stretch{1}}
\begin{minipage}[t]{.48\columnwidth}
\caption{(Color online)
Maxima of the spectral function $\rho$ at $T = 0.15\, T_c$ along the indicated constant $\omega$ slices.
The ratio of the Fermi velocities is $v_\perp/v_\parallel \simeq 13$.
The dashed line specifies the FS.
\label{F:cones}}
\end{minipage}
\vspace{-1.1em}
\end{figure}

In fig.~\ref{fig: density plot MDC} we plot the spectral function $\rho(\omega,\vk)$ along the $\omega=0$ plane.
As expected at low temperatures, the gap-generating mechanism discussed above reduces the surface of quasi-normal modes intersecting the $\omega=0$ plane to four nodal points. As depicted in fig.~\ref{F:cones}, close to the nodes and at small $|\omega/\mu q|>0$, the quasi-normal modes have the structure of asymmetric Dirac cones.
At the node, one can characterize the dispersion relation of the quasi-normal modes by the Fermi velocities $v_\bot = | \partial_{k_{\bot}} \omega|$ and $v_\parallel = | \partial_{k_{\parallel}} \omega |$. Here $k_{\bot(\parallel)}$ is the momentum in the direction perpendicular (parallel) to the FS at the node, and $\omega(k)$ specifies the location of the quasi-normal mode. We find that $v_\perp/ v_\parallel \simeq 13$.

To compute the size of the gap along the FS as a function of angle and temperature, we need first to identify the FS. Since we are working at non-zero temperature and since the spectral function is gapped, the FS is somewhat ill-defined. We 
define the Fermi momentum $k_F(\theta)$ as the momentum at which the gap is minimized, or as the momentum at which the spectral function is maximized if a gap does not exist. This definition is similar to the one in \eg{} \cite{Kanigel:2006}.
We find that $k_F$ is almost independent of $\theta$.  A curious feature of our system is that $k_F$ decreases (by roughly 20\%) as the temperature is lowered approaching $k_F \sim 0.11 \, q \mu$ at low temperatures.

\begin{figure}[t]
\includegraphics[height=5.6cm]{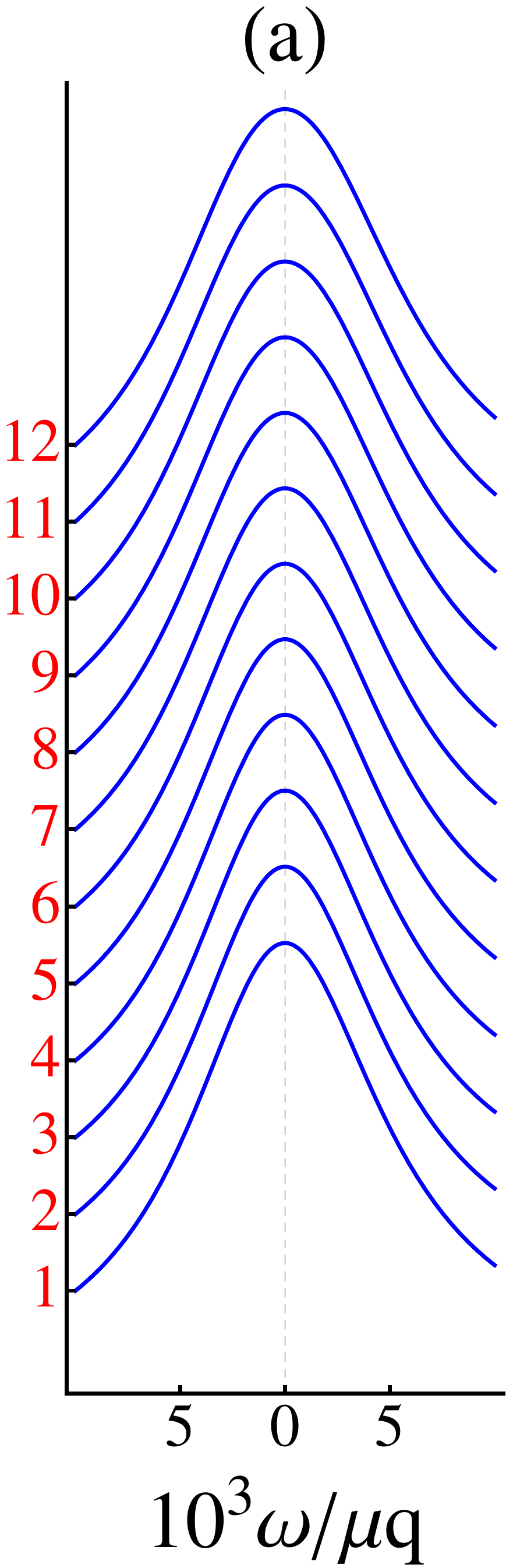}
\includegraphics[height=5.6cm]{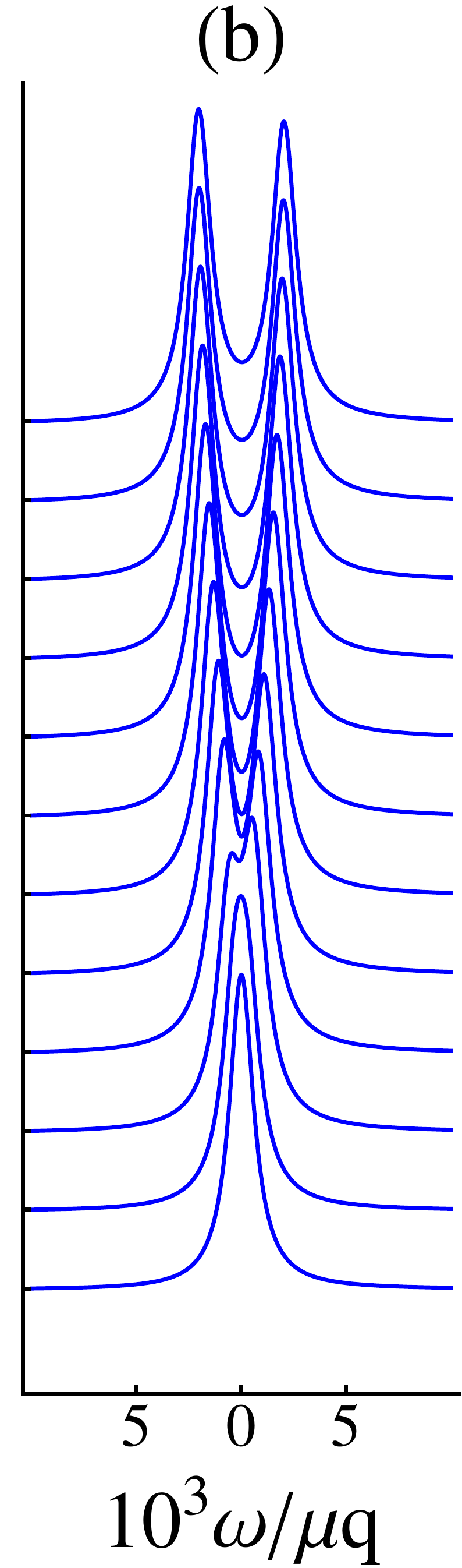}
\includegraphics[height=5.6cm]{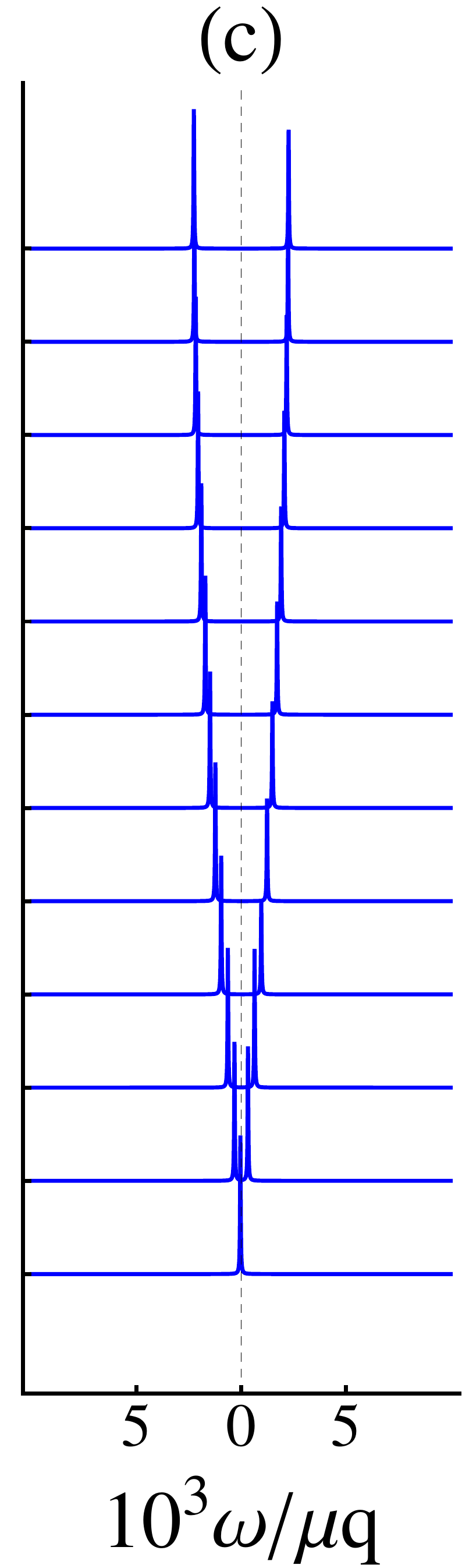}
\begin{minipage}[b]{0.35\columnwidth}
\includegraphics[width=\columnwidth]{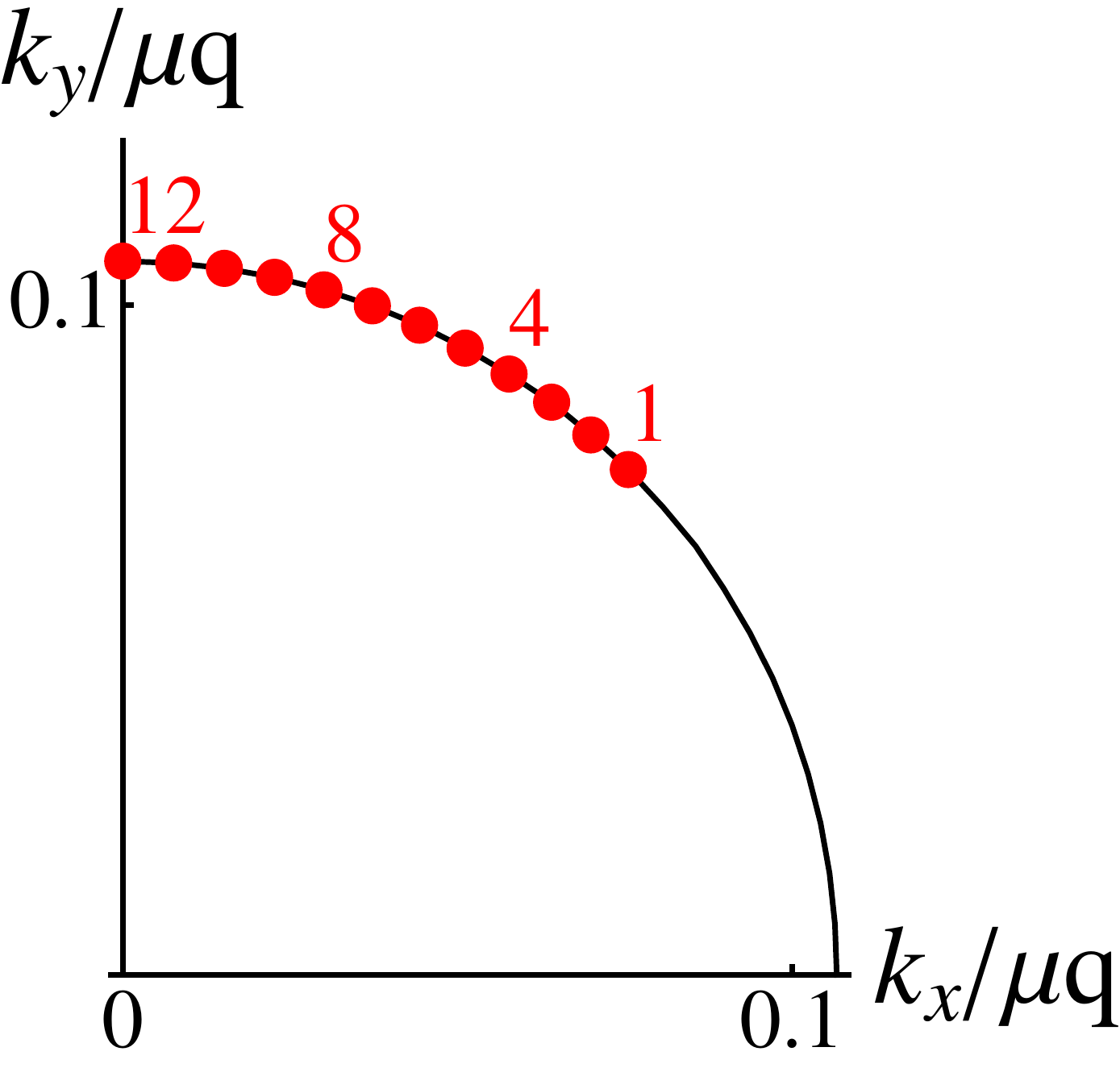} \\
\includegraphics[width=\columnwidth,trim=0 -40 0 0]{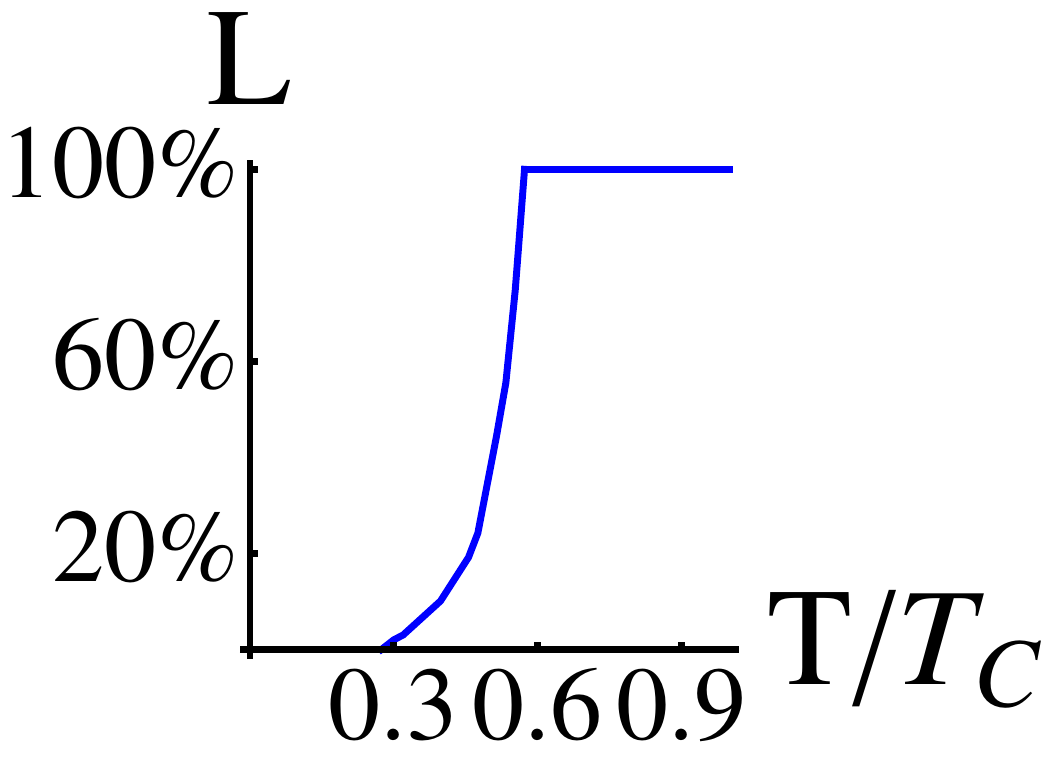}
\end{minipage}
\caption{(Color online) EDCs, $\rho(\omega, k_F(\theta),\theta)$, for several angles and temperatures. In the left panel, figures (a), (b) and (c) correspond to temperatures $T/T_c = 0.66,\,0.43,\,0.27$ respectively. The bottom right plot shows the dependence of the Fermi arc length on temperature. The FS at $T=0.43 \, T_c$ is plotted on the top right. \label{fig: EDCs at angles}}
\vspace{-.5em}
\end{figure}

The spectral function $\rho \big( \omega, k_F(\theta), \theta \big)$ is called the energy distribution curve (EDC). In fig.~\ref{fig: EDCs at angles} we plot the EDC for various angles, and for three different temperatures.
We define the gap $\Delta(\theta,T)$ as half the energy between the two maxima of the EDC. By definition, $\Delta=0$ when there is only one maximum.
At temperatures above $T_{\rm arc} =0.56 \pm 0.01 \, T_c$ thermal effects broaden the gap so that it is unobservable at all angles (fig.~\ref{fig: EDCs at angles}a).
For temperatures $T < T_{\rm gap} = 0.28 \pm 0.01 \, T_c $ (and with an angular resolution of $1/100$) a gap is observed everywhere except for the node located at $\theta=\pi/4$ (fig.~\ref{fig: EDCs at angles}c). In this case, $\Delta(\theta,T) = \Delta_0(T) |\cos(2\theta)|$
to very good accuracy as expected from mean field theory
\footnote{At first order, eigenvalue repulsion between degenerate states 
yields an energy gap proportional to the coupling.}. At intermediate temperatures the gap opens up only outside an interval $|\theta- \frac\pi4| \leq \theta_0(T)$ (fig.~\ref{fig: EDCs at angles}b). The interval where the spectral function remains ungapped is called a Fermi arc \cite{Norman:1998,Kanigel:2006,Kanigel:2007}. A plot of the dependence of the gap $\Delta$ on angle and temperature is shown in fig.~\ref{fig: gap vs angle}. The dependance of the arc length on temperature is given in fig.~\ref{fig: EDCs at angles}.

\textit{Discussion.} --
It is interesting to compare the holographic model discussed 
here with 
(ARPES) photoemission experiments on high $T_c$ superconductors
\cite{Damascelli:2003bi,Campuzano:2002,Zhou:2006}.
Such a comparison is problematic if one wants to get quantitative insight into the superconductive phase of the cuprates: Our holographic setup is merely a toy model for a true holographic $d$-wave superconductor with a well defined boundary theory; the fermions described here are not, a priori, related to the fermions forming the condensate; there is no underlying lattice in our model. Nevertheless it is thought-provoking that, even in the absence of several significant features of the cuprates, the holographic model manages to reproduce some of their notable properties.

One key feature of cuprate superconductors is the appearance of a $d$-wave gap in the superconductive phase \cite{Tsuei:2000}. The gap $\Delta$ is well fit by the function $|\cos k_x - \cos k_y|$ evaluated on the FS, as expected from a square lattice, and is essentially indistinguishable from the mean field result $\Delta = \Delta_0 \cos(2\theta)$.
The four nodes associated with such a gap and the Dirac cones emanating from them have been observed experimentally \cite{Mesot:1999}.
These features are similar to the ones we observe in our holographic setup, depicted in fig.~\ref{fig: gap vs angle} and \ref{F:cones}.
A typical experimental value of the ratio between the quasi-particle Fermi velocities is $v_\perp/v_\parallel \sim 20$ \cite{Campuzano:2002}. Such a ratio can be obtained in the holographic setup by reducing $\eta$.

Fermi arcs are observed in the pseudogap phase of high $T_c$ materials, at temperatures $T_c < T < T^*$ \cite{Norman:1998,Kanigel:2006,Kanigel:2007}. As described above, these are finite angular intervals along which the FS remains ungapped. The holographic Fermi arcs displayed in fig.~\ref{fig: density plot MDC} and \ref{fig: EDCs at angles} are distinct in that they appear in the superconductive phase, at temperatures $T_{\rm gap}<T<T_{\rm arc}<T_c$. In addition, the dependence of the arc length on the temperature
(fig.~\ref{fig: EDCs at angles}, lower right) is certainly nonlinear whereas in \cite{Kanigel:2006} it was observed that the arc length depends linearly on the temperature.


\begin{acknowledgments}

We thank D.~Huse, M.~Porrati, R.~Rahman, F.~Rocha and D.~Vegh for useful discussions.
AY is supported in part by the US DOE Grant No. DE-FG02-91ER40671 and by the
US NSF Grant No. PHY-0652782. CH and FB are supported in part by the
US NSF Grant No. PHY-0844827 and PHY-0756966.

\end{acknowledgments}

\bibliography{bibnoeprint}

\end{document}